\documentclass{revtex4} 
\usepackage{epsfig}
\usepackage{graphicx}
\begin{document}
\bibliographystyle{revtex}
\title{Associated neutralino-neutralino-photon  production at NLC}
\author{H.~Baer}
\email[]{baer@hep.fsu.edu,}
\affiliation{Physics Department, Florida State University, Tallahassee,
FL 32306-4350}
\author{A.~Belyaev}
\email[]{belyaev@hep.fsu.edu}
\affiliation{Physics Department, Florida State University, Tallahassee,
FL 32306-4350}
\date{\today}
\begin{abstract}
   We study the potential of an $e^+e^-$ linear collider to search for
   neutralino-neutralino-photon  production. Our analysis shows that this
   signal is not viable under realistic expectations for electron beam 
   polarization due to large Standard Model backgrounds. Such a search 
   would be possible only if beam polarizations of 
   near 100\% could be achieved.
   
 \end{abstract}
\maketitle
\label{sec:intro}
As part of the Snowmass effort to investigate the physics potential of
next linear colliders, we have studied the possibility of extending
the parameter space reach of $e^+e^-$ colliders by searching for
production of $\tilde\chi_1^0\tilde\chi_1^0\gamma$ events.
 
\begin{figure}[htb]
  \begin{minipage}[b]{.46\linewidth}
\noindent
\mbox{\epsfig{file=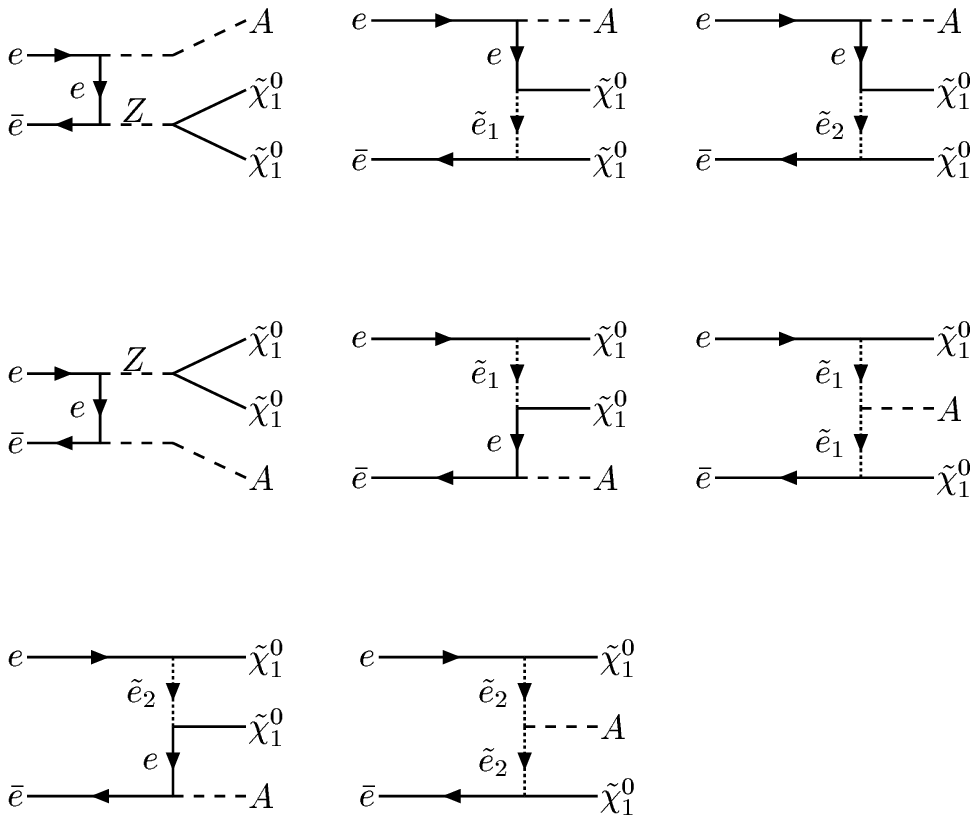,width=1\textwidth}}
\caption{\label{sign-diagrams} Feynman diagrams
for $e^+e-\to\tilde\chi^0_1\tilde\chi^0_1\gamma$ process}
\vspace*{0.2cm} 
  \end{minipage}
\hspace*{0.4cm}
  \begin{minipage}[b]{.46\linewidth}
\mbox{\epsfig{file=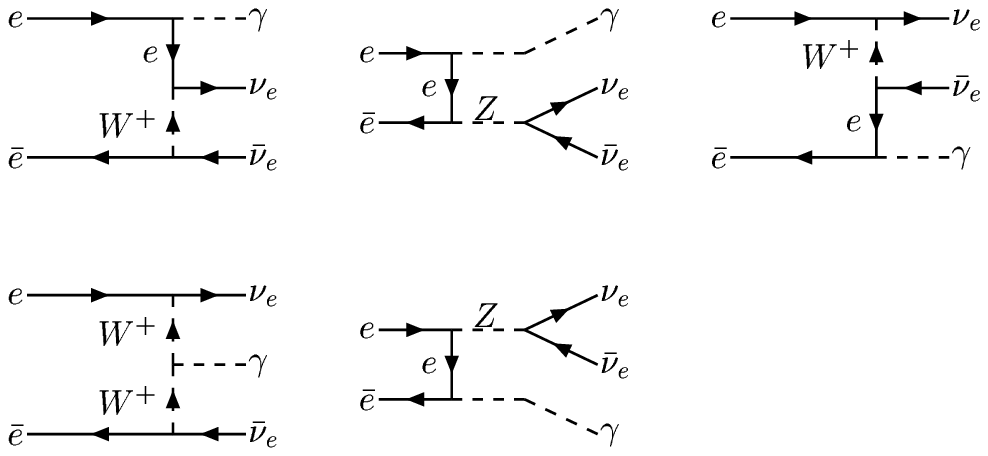,width=1\textwidth}}
\caption{\label{back-diagrams} Feynman diagrams
for $e^+e-\to\nu\nu\gamma$  SM background process}
\end{minipage}
\end{figure}

The photon plays a role of the trigger for pair production of massive 
invisible particles such as neutralinos in $e^+e^-$ collisions.
The Feynman diagrams for the $e^+e-\to\tilde\chi^0_1\tilde\chi^0_1\gamma$
process are shown in Fig.~\ref{sign-diagrams}. 
The main irreducible background for
this reaction  is the Standard Model (SM) process of neutrino pair plus photon
production --- $e^+e-\to\nu\nu\gamma$. The background Feynman diagrams 
are shown in  Fig.~\ref{back-diagrams} 
One can expect strong background dependence on the  polarization 
of the electrons in the initial state since $W$-boson exchange 
with pure  left(right)
coupling to the electron(positron) occurs. 
Those diagrams  could be "switched off" in the
ideal case case of the 100\% right polarized electron beam. If we imagine this
idealized situation, then the main SM background will come from the off-shell
Z-boson (on shell Z-boson contributions can be eliminated by using cut on the
photon energy).

To illustrate this situation qualitatively
we have chosen the center of  mass energy 
of $e^+e^-$ equal to 500 GeV as well as the following set of the
supersymmetric parameters:\\
1) gaugino masses $M_1$=$M_2$=200 GeV,\\
2) selectron masses: $m_{\tilde{e}_{1,2}}$=500 GeV,\\
3)$\tan\beta=3$ and $\mu=300$~GeV, \\
for which $m_{\tilde\chi^0}=157$~GeV.
We first apply a cut on the photon transverse momentum
$p_T^\gamma>10$~GeV to avoid soft and collinear divergences of the
cross section for this tree-level process. 
Our calculations has been done using the CompHEP software
package~\cite{Pukhov:1999gg}.

For this choice of parameters, in case of unpolarized electron-positron beams
the signal is 0.14 fb  while the background is 2300 fb i.e. 
about  4 orders of
magnitude bigger then the signal. In case of 100\% left-polarized  electron
(100\% right-polarized positron)
beam signal and background cross section increase by about a factor 3 and
become   0.36 and 7300 fb respectively. As expected, the situation for
S/B ratio drastically changes in case of opposite 
electron (positron)polarization. 
For a 100\% right-polarized electron beam 
the signal drops only by factor 2 compared to 100\% left-polarized case and
becomes equal to  18 fb while background is reduced by factor more than 20 
down to  342 fb.

\begin{figure}[htb]
\begin{minipage}[b]{.46\linewidth}
\mbox{\epsfig{file=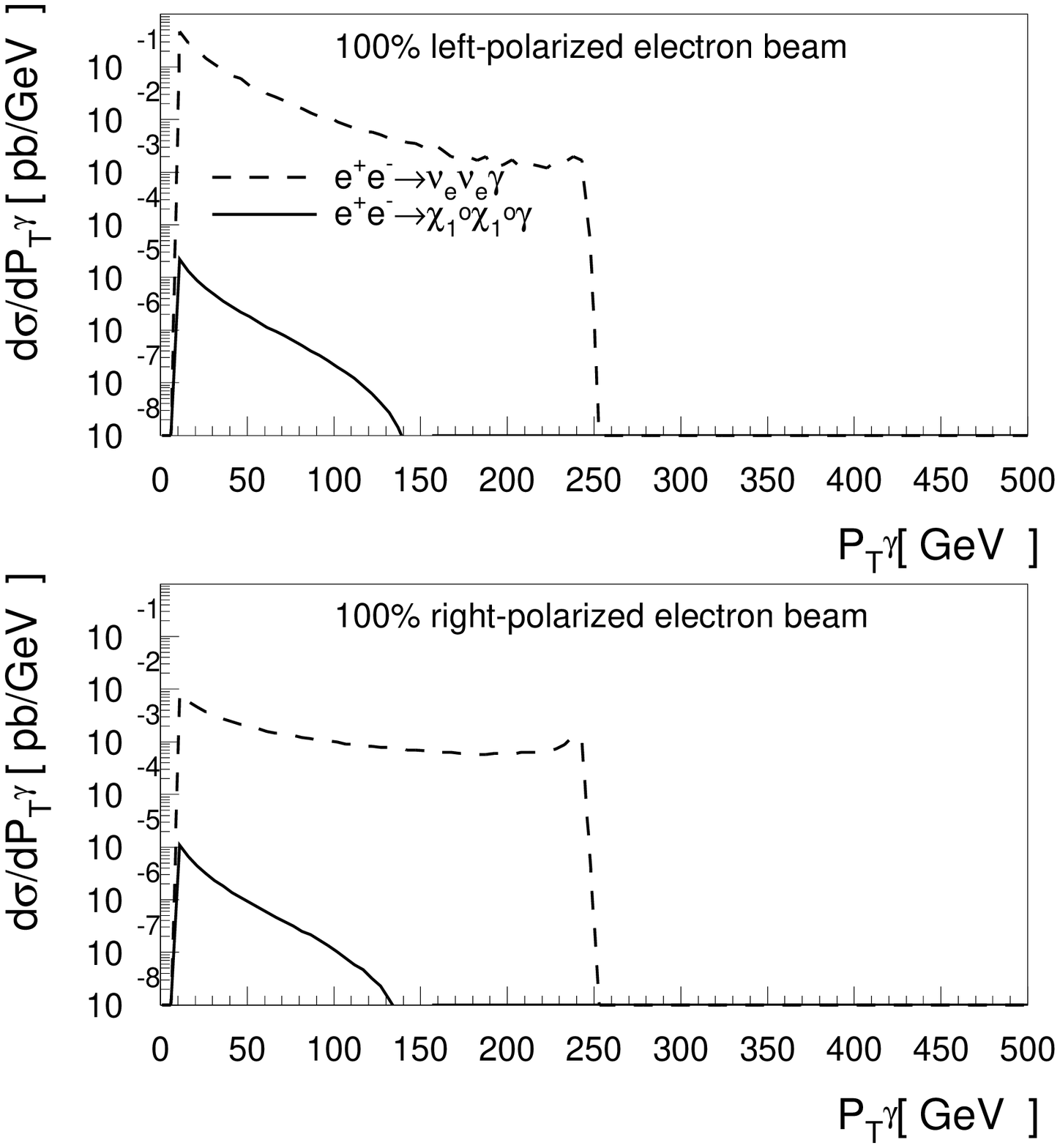,width=1\textwidth}}
\caption{\label{photon-pt} 
 Photon transverse momentum distribution
 of  $e^+e-\to\tilde\chi^0_1\tilde\chi^0_1\gamma$ signal process (solid line)
 and its $e^+e-\to\nu\nu\gamma$ background process (dashed line).
Upper plot presents distributions for  100\% left-polarized electron
while bottom plot is for  100\% right-polarized electron case.}
\end{minipage}
\hspace*{0.4cm}
\begin{minipage}[b]{.46\linewidth}
\mbox{\epsfig{file=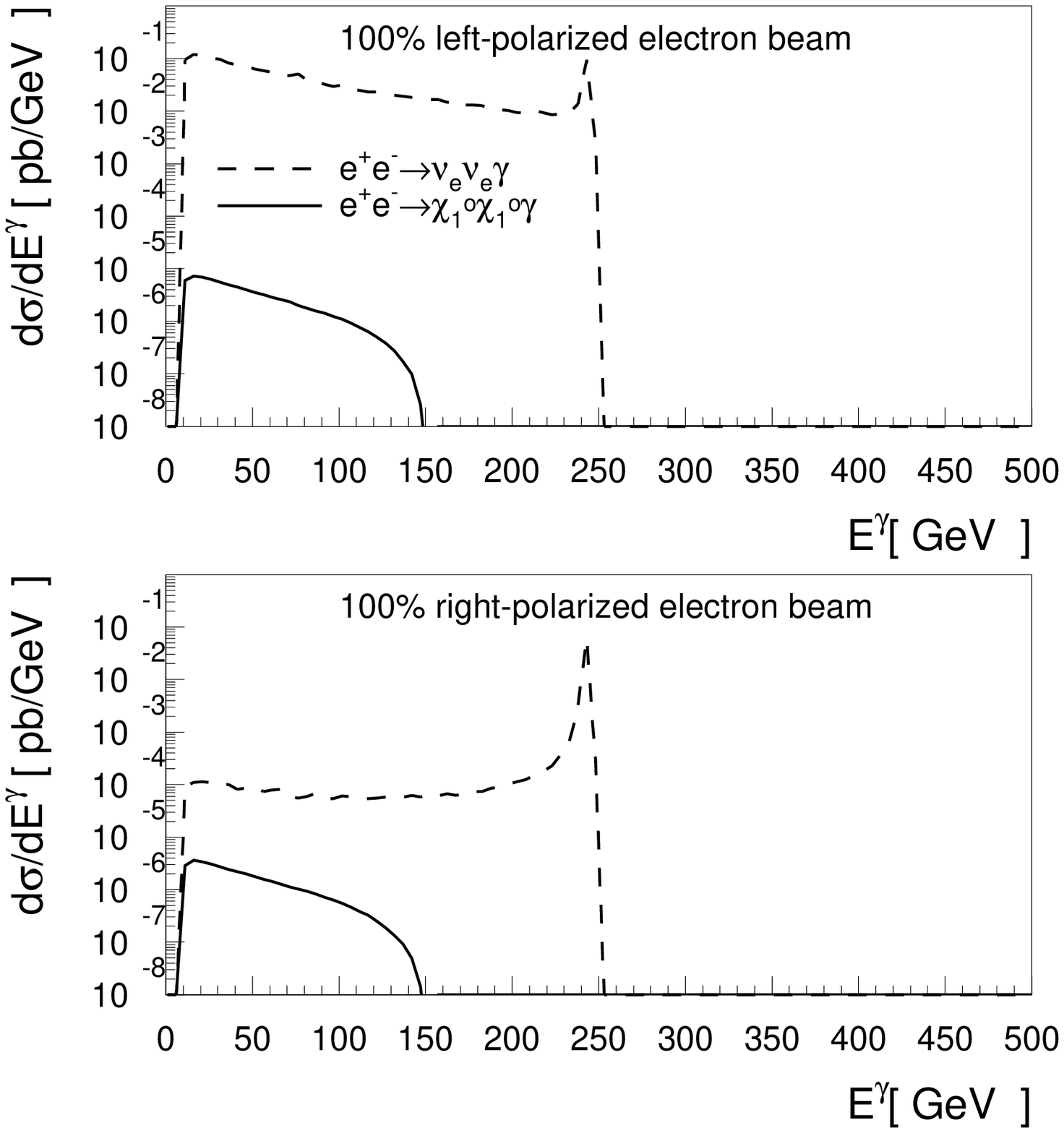,width=1\textwidth}}
\caption{\label{photon-e} Photon energy distribution
 of  $e^+e-\to\tilde\chi^0_1\tilde\chi^0_1\gamma$ signal process (solid line)
 and its $e^+e-\to\nu\nu\gamma$ background process (dashed line).
Upper plot presents distributions for  100\% left-polarized electron
while bottom plot is for  100\% right-polarized electron case.}
\vspace*{0.2cm} 
\end{minipage}
\end{figure}

Further improvement of the signal to background ratio could be done by
elimination of the on-shell Z-boson contribution for the background. To do
this one can study the transverse momentum and energy distribution  of the
photon and apply corresponding kinematical cuts. In
Figure~\ref{photon-pt},\ref{photon-e} we present photon transverse momentum and
energy respectively. The first natural cut following from
the $p_T^\gamma$ distribution is $p_T^\gamma<150$~GeV. The $E^\gamma$
distribution of the background has the bump around 240 GeV which is 
connected with the Z-boson resonance in $\nu\nu$ mass distribution. We
apply $E^\gamma<150$~GeV cut to eliminate  Z-boson on-shell contribution. 
Suggested $P_T^\gamma$ and $E^\gamma$ cuts reduce  background further by
factor more then 30.  It becomes now 10 fb. The signal remains
unchanged. The signal to background ratio is still $(0.18\ fb/10\ fb)\sim
1/50$. 

Finally we have applied detector acceptance cut for the photon
$|\cos\theta_\gamma|<0.95$ which reduces the signal only by 14\%
(down to 0.16~fb)  while background drops by 33\% down to 7.7~fb
because background photon tends to be little bit more forward-backward
then that from signal.

\begin{figure}[htb]
\begin{minipage}[b]{.46\linewidth}
\mbox{\epsfig{file=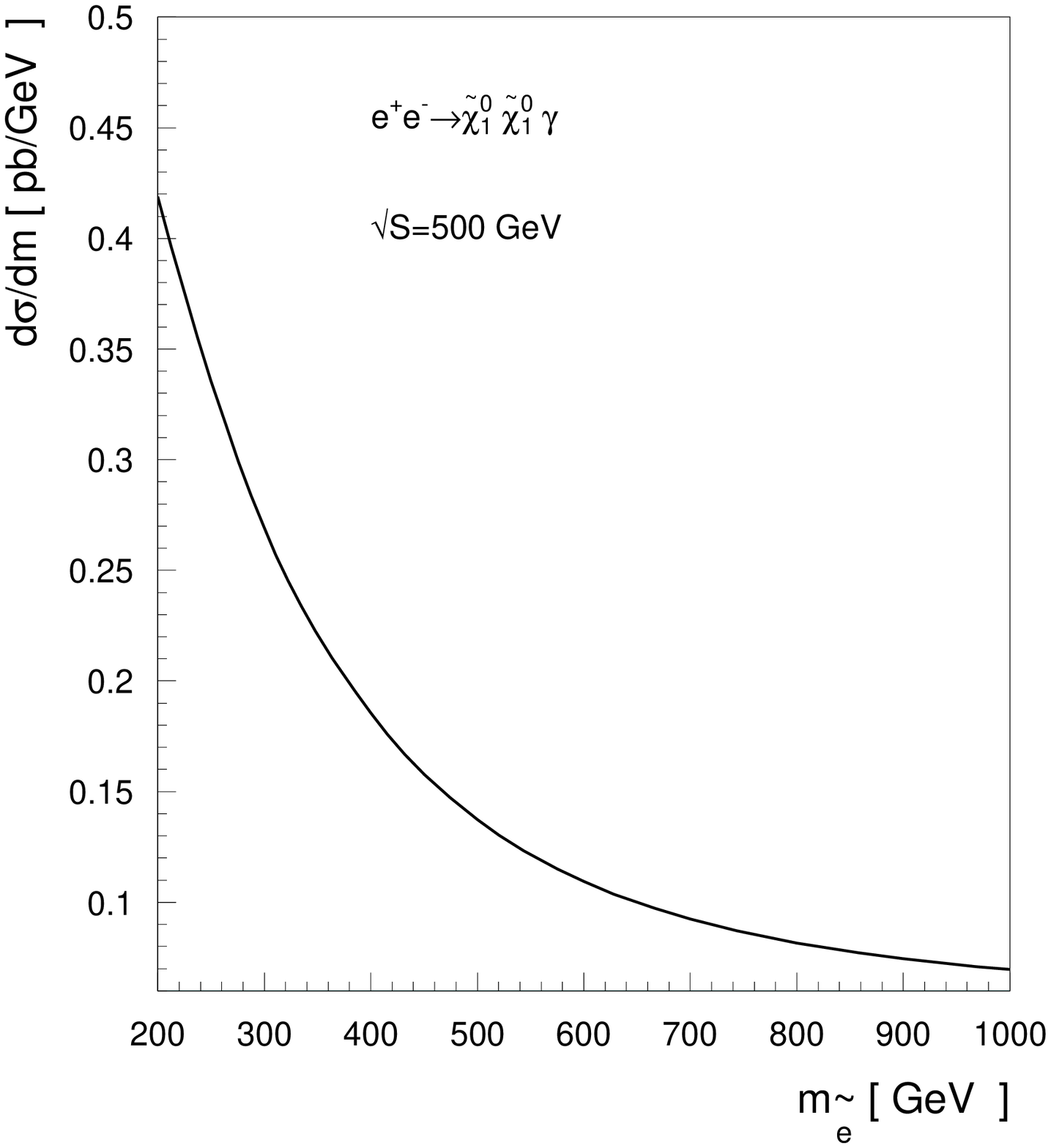,width=1\textwidth}}
\caption{\label{se} Cross section of $e^+e-\to\tilde\chi^0_1\tilde\chi^0_1\gamma$
as the function of the selectron mass for non-polarized electron beam.
SUSY parameters were taken as follows: $\mu=300$~GeV, $M_1=M_2=200$~GeV,
$\tan\beta=3$.}
\end{minipage}\hspace*{0.4cm}
\begin{minipage}[b]{.46\linewidth}
\mbox{\epsfig{file=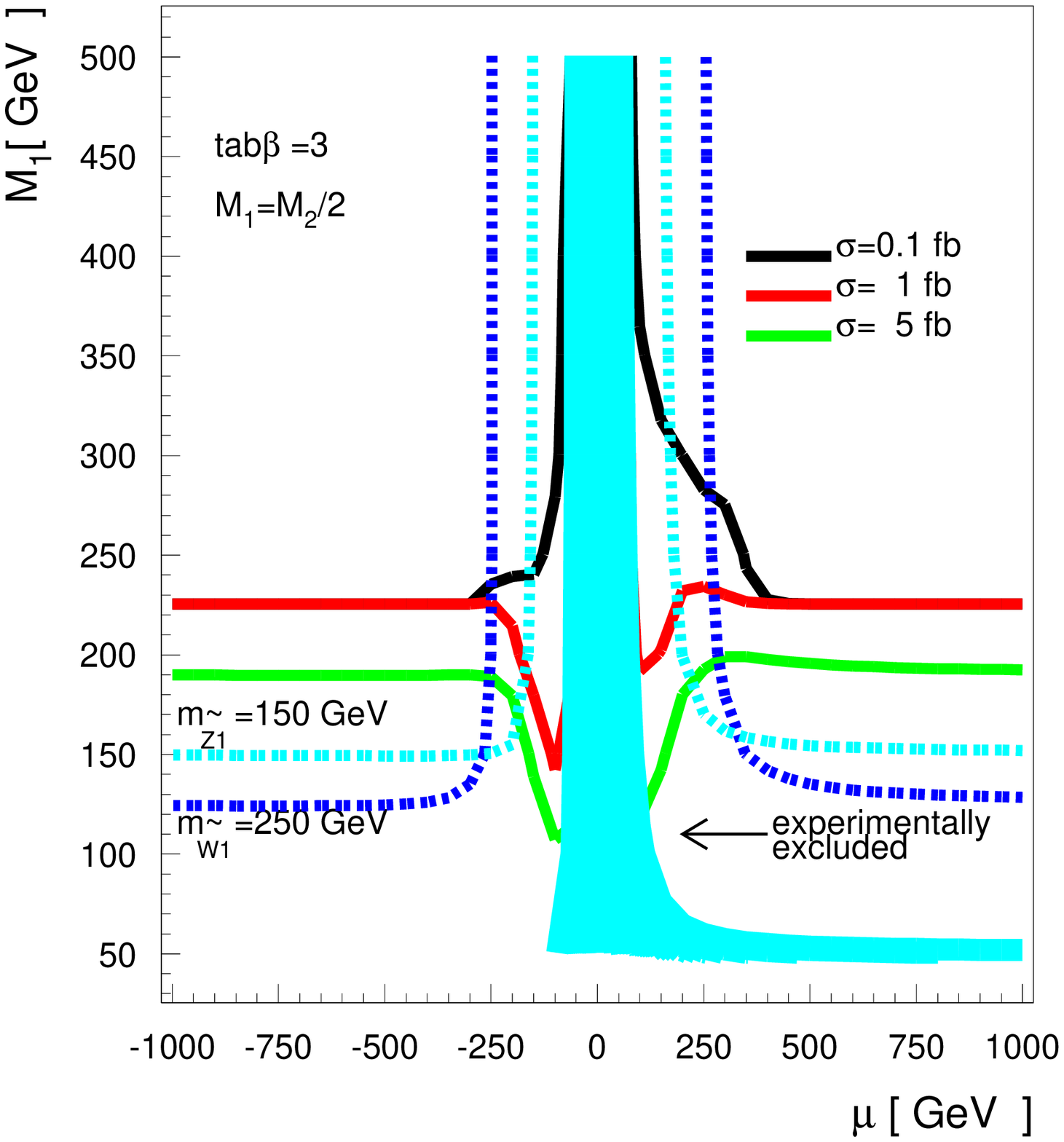,width=1\textwidth}}
\caption{\label{cs_level}Contour plot of the signal
cross section  and neutralino mass iso-levels in $(\mu, M_1)$ plane for
$m_{\tilde{e}_{1,2}}=500$~GeV.  }
\vspace*{0.2cm} 
\end{minipage}
\end{figure}

There is no any further cuts were found  to significantly reduce the
background. Therefore one could check  the possible signal enhancement
factors: signal dependence on selectron mass~(Fig.~\ref{se}) and
neutralino mass together with  neutralino mixing matrix elements. 

In
Fig.~\ref{cs_level} we present contour plots of the signal cross section 
and the neutralino mass (150 GeV) and chargino mass (250 GeV)
in the $(\mu, M_1)$ plane for which
$m_{\tilde{e}_{1,2}}$ was fixed to 250 GeV. For this figure we
use also $M_2=2 M_1$ and $\tan\beta =3$.

One can see from
Fig.~\ref{cs_level} that signal cross section could be as big as 5 fb for
$M_1 \leq 150$~GeV and $m_{\chi^0_1}\sim 150$~GeV. This happens when the
neutralino is almost a pure gaugino. Thus, the process looks viable for
pure right polarized beams.

Using a more realistic electron right handed polarization of 95\%, 
the background level after all cuts is at the 150 fb level, 
far beyond the best signal levels.
Therefore the process under study is not likely to be useful
for extending the SUSY parameter space reach of an $e^+e^-$ linear collider
unless an almost pure level of beam polarization can be achieved.

\begin{acknowledgments}
  The work of A.B.  is supported in part by the U.S. Department
  of Energy under contract No.~DE-FG02-97ER41022 .
\end{acknowledgments}
\bibliography{P3_Belyaev_036}
\end{document}